\documentclass{svproc}
 
\usepackage{minted}
\usepackage{graphicx} 
\begin{document}
 
\title{Embedding Quality Assurance in project-based learning -- An Experience Report}

\author{Maria Spichkova}
 \institute{School of Computing Technologies, RMIT University, Melbourne, Australia\\
\email{maria.spichkova@rmit.edu.au}\\ 
}
\maketitle              

\begin{abstract}
In this paper, we share our lessons learned from more than a decade of teaching software quality aspects within Software Engineering (SE) courses, where the focus is on Agile/Scrum settings: final year software development projects and the course on SE Project Management. Based on the lessons learned, we also provide a number of recommendations on embedding quality assurance topics in the project-based learning with Agile/Scrum context.
\end{abstract}

\section{Introduction}
Quality Assurance  (QA) is a vital part of the software development process. Academics and practitioners have elaborated on various approaches to measure and evaluate software quality, as well as to improve it based on the evaluation results. 
However, even effective approaches can help to improve software quality only if they are really used. Over the past decade of teaching software engineering courses, we have observed that many students struggle to understand the importance of quality assurance, particularly when working in Agile/Scrum settings. 
The experience report by Thompson and Edwards~\cite{thompson2000teach} highlighted that further investigation on the most appropriate mechanisms for
teaching quality, standards, and process improvement.  

Ardic and Zaidman~\cite{ardic2023hey} conducted a comprehensive analysis to investigate what sources (higher education, self-training, or training in industry) software engineers get their testing skills from. The analysis included a survey with software testing practitioners. Notably, the most popular option the respondents selected as their primary source of testing skill was `self-taught'. Moreover, many respondents expressed an opinion that they would prefer to have learned more about testing activities during their university studies. 
This emphasizes the importance of increased focus on QA activities within SE courses.

Marrington et al.~\cite{marrington2005quality} conducted a survey on students' perceptions regarding QA activities they conducted. Marrington et al. concluded that in Agile settings, it is especially difficult to ensure that the students invest enough time and effort in the QA activities. 
Our experience aligns with these conclusions. It might indeed be easier from a learning and teaching perspective to teach QA aspects on the basis of the Waterfall methodology. However, this would not really prepare students for the real industrial project. The agile approach became very popular over the last 20 years~\cite{hoda2018rise,hoda2017systematic}. More and more software development companies nowadays prefer to use the agile approach. As per recent State of Agile surveys, 97\% of respondents’ organisations adopted agile practices for at least some of their teams/projects - this holds for the surveys organised in 2018, 2021 and 2025~\cite{soAgile2018,soAgile2021,soAgile2025}.
One of the most successful methodologies to support Agile software development is currently Scrum~\cite{schwaber2011scrum}.  

Integrating capstone projects into the curriculum provides students with real industry experience~\cite{balaban2018software,bastarrica2017can,knudson2018global,ICSE_2018_capstone}. However, the course structure and assessments should be well-developed to focus students on critical aspects and ensure that they gain the most from this learning opportunity. 
Many studies highlight that capstone projects might be especially effective when the projects are (1) based on real-life problems, and (2) conducted with real clients / industrial partners, see, e.g., \cite{bruegge2015software,simic2016enhancing,daun2016project}.
Involving industrial partners in the project-based learning is also useful to ensure the authenticity of the  assessment tasks~\cite{redecker2013changing,dos2016pbl}. 
However, it's critical to ensure that the industrial partners are aware of the details of the course structure and assessments, as well as of the expected Course Learning Outcomes (CLOs). Thus, a close collaboration with the course coordinator and/or academic mentors of the projects is necessary.

In this experience report, we share our lessons learned from teaching of software quality aspects within Software Engineering (SE) courses, where the focus is on Agile/Scrum settings.   

\section{Background: QA in Agile/Scrum}
\label{sec:background}

 In a Scrum project, the development team is typically small, up to 10, more commonly 6-8 people. The team organises their work in so-called \emph{sprints}, which are fixed-length phases (typically 1 or 2 weeks long, in some cases up to 4 weeks). At the very beginning of each sprint, the team holds \emph{sprint planning meeting}, where a plan for the sprint is created. During a sprint, the team should have daily stand-ups (short meetings, where the team members update each other on the latest status), and at the end of the sprint the completed items are delivered and presented to the key stakeholders in the \emph{sprint review meeting}, after which an internal \emph{sprint retrospective (retro) meeting}  is organised to discuss ways to increase quality and effectiveness. 
 From what we observed, many students struggle with Agile/Scrum idea that an item can be counted as completed only when it satisfies all quality measures required for the product.

One of the core artefacts of Agile/Scrum is \emph{Product backlog}, the list of known requirements. For each item in the product backlog (so-called \emph{Product Backlog Item}, PBI), there should be specified corresponding priorities, associated expected efforts, and status (e.g., ``to-do", ``in-progress", ``done"). The requirements are typically written as \emph{user stories}, i.e. in the format\\
~\\
 \emph{As a $<$type of user$>$ \\ I want to $<$goal action$>$\\  so that $<$goal reason/benefit$>$ }

All core quality measures required for the product have to be specified as so-called \emph{Definition of Done} (DoD). A PBI can be counted as completed and labelled as ``done" only if it's fully done in terms of DoD. 
The content of DoD depends on the nature of the project. A systematic review on the use of Definition of Done on agile software development projects was presented by Silva et al.~\cite{silva2017systematic}.

According to the recent State of Agile~\cite{soAgile2021} (the latest version doesn't present statistics on this particular topic, providing more emphasis on the use of AI approaches), the following quality assurance techniques and tools have been used by respondents:
\begin{itemize}
    \item 54\% applied automated build tools,
    \item 54\% used  automated unit testing,
    \item 53\% used continuous integration,
    \item 47\% applied release/deployment automation tools,
    \item 37\% used static analysis of the developed application/system,
    \item 35\% used automated acceptance testing.
\end{itemize}
In a survey conducted by Diebold et al. \cite{diebold2015practitioners}, similar results have been identified: 
\begin{itemize}
    \item all 10 interviewees (100\%) described the usage of automated tests in their companies, which are usually part of a continuous integration and nightly builds,
    \item 40\%  explicitly mention additional code reviews and automated static analysis
    \item 40\% explicitly mentioned manual tests,
    \item 10\% (only one one company) mentioned that they also conduct reviews of each User Story they define. 
\end{itemize}

\section{Lessons Learned}

The lessons learned that we discussed in this report are based on our experience of teaching and supervision of team-based projects conducted using Scrum methodology. 
Overall it is based on teaching of three types of courses: capstone projects and the SE Project Management (SEPM) course.

Since 2014, we have supervised a number of so-called \emph{capstone projects} each year, i.e., final-year courses that aim to provide the cumulative experience of a study program. These projects have been provided within several courses provided for both Bachelor and Master students at their final year of study, within the following study programs:
\begin{itemize}
    \item Bachelor of IT (BIT),
    \item Bachelor of Software Engineering (BSE),
    \item Master of Computer Science (MSC),
    \item Master of IT (MIT).     
\end{itemize}
Since 2019, we have also coordinated teams within the SEPM course, where the project-based assignment focuses on creating Agile/Scrum artefacts, including product and sprint backlogs, user story cards with detailed acceptance criteria, and notes for planning and retrospective meetings. 
Since 2019, we have worked on continuous improvement of the SEPM course to help students overcome the above issues. The summary of the results is presented in~\cite{SPICHKOVA2025agile}. We also re-designed these courses~\cite{spichkova2019industry} to cover the E$^3$:
Engagement, Experience, Employability.  However, this above summary and the re-desing don't cover any QA aspects. 
Most of the lessons learned within the process of re-design were related to the course and program structure as well as to students' preferences and communication aspects (both within the team and with clients/industrial partners). In this report, we focus on teaching and learning of Quality Assurance aspects. 

The recommendations we provide below differ by the type of the course. Some solutions might be a good fit for the team-based capstone projects only, where the whole set of activities within the course is conducted with a software development project. Other solutions might work well for the courses, where tasks are provided as separate assignments with scenarios unrelated to each other (which is the case of the SEPM course).

~\\
\textbf{Lesson Learned 1:}
We have observed that it's hard for students to cope with the fact that the system requirements, design and  architecture might evolve over the project. 
It is often the case that the assessment tasks are typically well-formulated by instructors, and only minor clarifications are needed (if any). It's a rare exception when an instructor changes some parts of an assessment task. Therefore, after completing school and university studies, novice software developers often have the perception that tasks will be provided to them as static and well-formulated requirements, and nothing will change throughout the project.  
We also identified that despite prior completion of the SEPM course (which has a strong focus on Agile/ Scrum), many students tend to think about software development in a pure waterfall manner, and struggle to deal with the agile approach, especially in terms of sprint planning and quality assurance. 

\textbf{Recommendation 1a: Capstone projects.} It might be useful to `embed'  in the capstone project the case of requirements or scope change. This might be discussed with the industrial partners in advance, to clarify that reasonable adjustments are welcome for the projects. 
In the seldom case where the industrial partners are perfectly prepared by having a well-defined scope already specified (or even over-prepared by having a detailed list of functional requirements specified by them in advance), the course coordinator might propose sharing with the team at the very beginning of the project an incomplete description and refining it later, to provide a more realistic experience.

\textbf{Recommendation 1b: Courses with separate assignments.} 
The assignments might be based on the scenarios that highlight the possibility of changes/adjustments in the scope or of the particular requirements. For example, in the SEPM couse, we include in the scenario to be used for the burndown chart creation the following element:\\
\emph{``At the sprint planning meeting for Sprint 2, the Product Owner requested to add to the Product Backlog two low-priority items: PBI-42 and PBI-43. The team and Scrum Master agreed to this change. The team estimated the efforts for them as 13 story points each. 
At that meeting, the Product Owner decided to remove PBI-24 and PBI-25 from the Product Backlog (previously, PBI-24 was estimated as 8 story points, and PBI-25 was estimated as 5 story points). Both items have been previously assessed as medium-priority items. The team and Scrum Master agreed to this change as well.''}

~\\
\textbf{Lesson Learned 2:}
We have observed that it's challenging for students to understand the importance of eliciting and properly specifying requirements by the team.

\textbf{Recommendation 2a: Capstone projects.} 
The course coordinator should check the materials to be provided by the industrial partners to the teams. 
In the (seldom) case the industrial partners are over-prepared by having a detailed list of functional requirements specified by them in advance, this list shouldn't be provided to the students directly. The teams should have an opportunity to facilitate requirements engineering sessions to specify the set of requirements by themselves. 

\textbf{Recommendation 2b: Courses with separate assignments.} 
A simulation of requirements elicitation activities might be provided. However, it should be taken into account by the course coordinator that this type of assignment might be very time-consuming to prepare.

~\\
\textbf{Lesson Learned 3:} 
Some students miss the connection between requirements specification and system testing, considering quality of the code and corresponding testing activities as a `nice-to-have' elements that should be added only if time allows.

\textbf{Recommendation 3a: Capstone projects.} 
The following artefacts should be considered as a compulsory part of the project deliverables:
 (1) Specification of the test cases for system testing; (2) Unit tests providing the code coverage of a minimum of 70\% for achieving the pass/credit grade.  
 It would make sense to include these considerations on the course level and specify them explicitly in the assignment description and, if applicable, in the assignment rubrics (for the definition and examples of rubrics, see~\cite{mccauley2003rubrics}). 

 \textbf{Recommendation 3b: Courses with separate assignments.} 
 It may be useful to include tasks in the Acceptance Criteria (AC) specification, as well as in the test case specification for system testing, based on the provided AC. 

~\\
 In our previous work, we proposed including research components in capstone projects as a bonus task, see~\cite{spichkova2017autonomous}. The initial idea demonstrated that the teams of well-motivated students might not only learn core research method concepts but also produce great results even within very short, i.e., one/two weeks long, components (see, e.g., \cite{christianto2018enhancing,clunne2017modelling,sun2018software,chugh2019automated,gaikwad2019voice,spichkova2020gosecure}). However, we also observed that this indeed is valid only for the case when the members of the team are well-motivated and keen to learn new knowledge. 

~\\
\textbf{Lesson Learned 4:} 
Embedding research components in capstone projects is beneficial only for well-motivated and top-performing students who are keen to learn new knowledge. Lower-performing students tend to perform poorly in short research components, particularly when compared to their performance in development tasks. Most students attempt to do their best within the final year capstone project. For lower-performing students, focusing on tasks they are already familiar with from previous courses might be beneficial. 
Including research components means that students will work on novel topics and learn additional skills, which may be overwhelming for some students if the component is to be completed concurrently with the software development tasks. In this case, students might decide to take shortcuts in QA tasks while being under time pressure.  

\textbf{Recommendation 4a: Capstone projects.} 
We suggest including research components only as an additional, bonus task, selected only by the teams/students who are keen to gain the research-related experience.

\section{Summary}

In this experience report, we share our lessons learned from over a decade of teaching software quality aspects in Software Engineering (SE) courses that focus on Agile and Scrum methodologies. 
Based on the lessons learned, we provide several recommendations for embedding quality assurance topics in project-based learning within an Agile/Scrum context.


\end{document}